\documentstyle[12pt]{article}
\begin{document}
\large
\renewcommand {\baselinestretch} {1.0}
\setcounter {page} {1}
\begin{center}
{\bf Some Consequences of the Law of Local Energy Conservation
in the Gravitational Field}\\
\vspace{0.3cm}
\par
Kh.M. Beshtoev
\vspace{0.2 cm}
\par
Joint Institute for Nuclear Research, Joliot Curie 6, 141980 Dubna,
\par
Moscow region, Russia\\
\end{center}
\par
{\bf Abstract}\\
\par
At gravitational interactions of bodies and particles  there appears the
defect of masses, i.e. the energy yields since the bodies
(or particles) are attracted. It is shown that this changing of the
effective mass of the body (or the particle) in the external gravitational
field leads to changes the measurement units: velocity and length (relative to
the standard  measurement units). The expression describing the
advance of the perihelion of the planet (the Mercury) has been obtained.
This expression is mathematically identical to Einstein's equation for the
advance of the perihelion of the Mercury.

\section{Introduction}

At  gravitational interactions there appears a defect masses [1], i.e.
some an energy yields since the bodies (or particles) are attracted.
In the previous work [1] it was shown, that the radiation spectrum
(or energy levels) of atoms (or nuclei) in
the gravitational field has a red shift since the effective mass of
radiating electrons (or nucleons) changes in this field.
This red shift is equal to the red shift of the radiation spectrum
in the gravitational
field measured in existing experiments. The same shift must arise when
the photon (or $ \gamma $ quantum) is passing through the gravitational
field if it participates in gravitational interactions.
The absence of the double effect in the experiments
means that photons (or $ \gamma $ quanta) are passing through the
gravitational field without interactions.
\par
This work is devoted to search for the influence of gravitational interaction
on the body (or particle) characteristics.

\section{\bf Some Consequences of the Law of Local Energy Conservation
in Gravitational Field}

a). We will consider the influence of the external gravitational field
$\varphi = -G\frac{M}{r}$ ($M$ is a mass of the external body, $r$
is a distance and $G$ is a gravitational constant) on
characteristics of a body (or particle) having a small velocity. The law of
local energy conservation in the classical case is presented in the
following form:
$$
E = \frac{m v^2_1}{2} + m \varphi_1 = \frac{m v^2_2}{2} + m \varphi_2
\eqno(1)
$$
or
$$
\frac{m (v_1^2 - v^2_2)}{2} = m (\varphi_2 - \varphi_1)
\eqno(2)
$$
The Eqs. (1) and (2) characterize the balance between kinetic and potential
energies (the smaller one energy, the bigger another energy and back). It is
necessary to take into account the fact that in equations (1), (2) the mass $m$
is included as a factor. Formally we can delete this factor and then we came
to the following senseless equation:
$$
\frac{ (v_1^2 - v^2_2)}{2} =  (\varphi_2 - \varphi_1)
\eqno(2)
$$
In a more strict form the law of local energy conservation can be rewritten
in the form:
$$
E = m c^2 + \frac{m v^2_1}{2} + m \varphi_1 =
    m c^2 + \frac{m v^2_2}{2} + m \varphi_2 .
\eqno(3)
$$
Then the Eq. (3) can be rewritten in th following form:
$$
E = m c^2 (1 + \frac{\varphi_1}{c^2}) + \frac{m v^2_1}{2}  =
    m c^2 (1 + \frac{\varphi_2}{c^2}) + \frac{m v^2_2}{2} .
\eqno(4)
$$
After introduction the new masses:

$$
m' = m (1 + \frac{\varphi_1}{c^2}) \qquad
m'' = m (1 + \frac{\varphi_2}{c^2})          ,
\eqno(5)
$$
and new velocities:
$$
v_1'^2 = \frac{v^2_1}{(1 + \frac{\varphi_1}{c^2})} \qquad
v_2'^2 = \frac{v^2_2}{(1 + \frac{\varphi_2}{c^2})} .
\eqno(5')
$$
Eq. (4) acquires the following form:
$$
E = m' c^2 + \frac{m' v'^2_1}{2}  =
    m'' c^2 + \frac{m'' v'^2_2}{2} .
\eqno(6)
$$
The eq. (6) means that in an external gravitational field the effective mass
body (or particle) changes. For clarification of this question, let us
consider a body (or particle) with mass $m$ in the external gravitational
field $\varphi$ in point $r$ and write the law of local energy conservation
for this system
$$
m c^2 = E = m c^2 + \frac{m v^2}{2} + m \varphi_ =
    m' c^2 + \frac{m' v'^2}{2}  ,
\eqno(7)
$$
where
$$
m' = m(1 + \frac{\varphi}{c^2}) \qquad \Delta m = m - m' =
m \frac{\varphi}{c^2}, \qquad or \qquad \frac{\Delta m}{m} =
\frac{\varphi}{c^2} ,
$$
and
$$
\Delta v^2 = v^2 - v'^2 = - v^2 (\frac{\varphi}{c^2}) \qquad or \qquad
\frac{\Delta v^2}{v^2} = \frac{\varphi}{c^2} .
\eqno(8)
$$
Eqs. (7), (8) mean that changing of mass ($\Delta m c^2$) of the body
(or particle) in the external gravitational field goes on kinetic energy
of this body (or particle).
\par
And what is the result we have come to? In contrast to classical physics,
the velocity of the body (or particle) is $v'$ and
$$
\Delta v^2 = v^2 - v'^2 = -v^2 \frac{\varphi}{c^2}, \qquad
\Delta v = - v \frac{\varphi}{2 c^2} .
\eqno(9)
$$
The real velocity of the body (or particle) differs on $\Delta v$ from
the classical one. This difference can be measured by using  Dopller
and M${\ddot o}$sbauer effects at moving the radiation atoms in the
gravitational field of the Earth. We also can measure this effect at
rotating planet around the star (the Sun). The real velocity of the rotating
planet will differ from the classical one of the rotating planet.
\par
b) Since the difference between the real and classical velocities
$\Delta v$ is very small, then this displace is very small. So, we can
see this effect at multiple rotation of the planet around the star (the Sun).
From equation (8) we see that this difference is bigger if
the value $\mid \frac{\varphi}{c^2} \mid$ bigger is. For the Sun system
the biggest effect will be obtained for the Mercury since its
distance from the Sun is the smallest (this effect is known as advance of
perihelia of the Mercury).
\par
Let come to the computation of this effect.
\par
The rotation period of a planet around a star (the Sun) on a nearly circle
orbit is given by the following expression [2]:
$$
P =2 \pi \sqrt{\frac{r^3}{G M}}                   ,
\eqno(10)
$$
where $G$ is a gravitational constant and $M$ is the Sun mass.
\par
If the circular velocity of the planet (the Mercury) is $v$ and it does
one circle for period $P$, then perimeter $D$ is
$$
D = P v                   .
\eqno(11)
$$
If the real
velocity of the Mercury is $v'$ and $p'$ is the real period of its rotation,
then the real distance $D'$ is
$$ D' = P' v'
\eqno(11)
$$
Let us introduce the following definition :
$$
\epsilon = \frac{G M}{c^2 r} = - \frac{\varphi}{c^2}
\eqno(12)
$$
The centrifugal force is equal to the attractive force on the absolute
value at the rotating planet around the star (the Sun) and therefore
$$
\frac{m v^2}{r} = G \frac{m M}{r^2}               .
\eqno(13)
$$
From Eq. (13) we obtain the following correlation:
$$
v = \sqrt{\frac{G M}{r}}       ,
\eqno(14)
$$
then the velocity variation $\Delta v$ is
$$
\Delta v = -\frac{1}{2} \frac{\sqrt{G M}}{r^{1.5}} \Delta r ,
$$
then using eq. (14) we obtain the relation
$$
\frac{\Delta v}{v} = - \frac{1}{2} \frac{\Delta r}{r}
\eqno(15)
$$
\par
Since the time measurement unit is fulfilled by the atomic
clock, then the decrease of
mass results in the increase of length (Bohr orbit [1, 3]),
then the decrease of the planet mass results also in increasing
the distances (or standard measure unit).
$$
r \sim \frac{1}{m}, \qquad m \rightarrow m - \Delta m (\Delta m > 0)
$$
$$
r' = r + \Delta r = r(1 + \frac{\Delta r}{r})
$$
$$
\frac{1}{m'} =
\frac{1}{m - \Delta m} \simeq \frac{1}{m} (1 + \frac{\Delta m}{m})
= \frac{1}{m}(1 + \epsilon)
\eqno(16)
$$
So, we obtain
$$
\frac{\Delta r}{r} = \epsilon
\eqno(17)
$$
\par
Besides we must take into account the relativistic factor $\gamma = \frac{1}
{\sqrt{1 - v^2/c^2}}$, since the planet is moving around the star (the Sun)
with velocity $v$
$$
\frac{1}{\sqrt{1 - v^2/c^2}} \simeq 1 + \frac{v^2}{2 c^2} = 1 + \delta  .
\eqno(18)
$$
Putting the eq. (14) in eq. (18) we get
$$
\delta = \frac{v^2}{2 c^2}  = \frac{\epsilon}{2} .
\eqno(19)
$$
Collecting eqs (17), (19) and using eq. (12) we obtain the following
equation:
$$
\frac{\Delta r}{r} = \frac{\epsilon}{2} + \epsilon = 1.5 \epsilon  ,
\eqno(20)
$$
or
$$
r' = r (1 + 1.5 \epsilon) .
$$
Putting eq. (20) in eq. (10), we come to the following expression for
the period of the planet:
$$
P' = P (1 + 2.25 \epsilon)  .
\eqno(21)
$$
\par
Using eqs (15) and (20), we obtain the following expression for
the velocity displace (variation):
$$
\frac{\Delta v}{v} = - \frac{1}{2} \frac{\Delta r}{r} = - \frac{1.5}{2}
\epsilon .
\eqno(22)
$$
\par
The Kepler's law states that the smaller the distance from the Sun,
the more the planet velocity is, i.e.
$$
\frac{\Delta v}{v} = \frac{v - v'}{v} < 0 .
$$
It takes place since the length measure unit grosses up
and, correspondingly, the velocity must gross on the absolute value. If to
take into account the Kepler's law, then
$$
\frac{\Delta v}{v} = + \frac{1}{2} \frac {\Delta r}{r} = \frac{1.5
\epsilon}{2} = 0.75 \epsilon
\eqno(23)
$$
or
$$
v' = v (1 + 0.75 \epsilon)
\eqno(23')
$$
Putting equations (21) and (23') in eq. (11), we obtain
$$
D' = P' v' = P v (1 + 2.25 \epsilon)(1 + 0.75 \epsilon) \simeq
P v (1 + 3 \epsilon) = D (1 + 3 \epsilon)
\eqno(24)
$$
or
$$
D' = D (1 + \frac{3 G M}{ c^2 r})
\eqno(24')
$$
Equation (24') as function of angle $\varphi$ has the following form:
$$
\varphi' = \varphi  + \Delta \varphi, \qquad
\Delta \varphi = 2 \pi 3 \epsilon = 6
\pi \frac{G M}{c^2 r}   ,
\eqno(25)
$$
for elliptic orbit $r \rightarrow r (1 - e^2)$ ($e$ is a larger
eccentricity and $\Delta \varphi$ is
$$
\Delta \varphi = \frac{6 \pi G M}{c^2 r (1 - e^2)}
\eqno(26)
$$
Eqs. (25), (26) are mathematically identical to Einstein's equation
[4]. These equations are obtained in the flat space by taking into
account changing of the effective mass of the planet in the external
gravitational field, i.e. by using the law of local energy conservation.
A more detailed consideration of this problem is presented in work [3].

\section{Conclusion}

So, the effective mass $m'$ of the planet rotation around the star
(the Sun) with mass $M$ on distance $r$ is
$$
m' = m (1 - \frac{\varphi}{c^2}), \qquad \varphi = G \frac{M}{r} ,
$$
and the corresponding connection energy is
$$
\Delta E = m \mid \varphi \mid = \Delta m c^2 .
$$
Half of this energy is the energy of the planet rotation around the
star (the Sun)
$$
m' v^2 = \mid G\frac{m' M}{r} \mid = \frac{m'}{2} \mid \varphi \mid,
\qquad v^2  = \varphi,
$$
$$
E_{kin} = \frac{m'v^2}{2} = \frac{m' \varphi}{2} ,
$$
and another half-
$$
E_{con} = \frac{m' \varphi}{2}
$$
is the energy lost of planet rotating around the star (the Sun).
\par
At gravitational interactions of bodies and particles  there appears the
defect of masses, i.e. the energy yields since the bodies
(or particles) are attracted. It was shown that this changing of the
effective mass of the body (or the particle) in the external gravitational
field leads to changes the measurement units: velocity and length (relative to
the standard  measurement units). The expression describing the
advance of the perihelion of the planet (the Mercury) has been obtained.
This expression is mathematically identical to Einstein's equation for the
advance of the perihelion of the Mercury.\\

\par
{\bf References}

\par
\noindent
1. Kh.M. Beshtoev, JINR Commun. P4-2000-45, Dubna, 2000;
\par
quanta-ph/0004074.
\par
\noindent
2. P.I. Bakulin, E.V. Kononovich, V.I. Moroz, Course of
\par
General Astronomy, M., Nauka, 1977.
\par
\noindent
3. P. Marmet, Physics Essays, 1999, v.3, p.468.
\par
\noindent
4. A. Einstein, Ann. Phys., 1911, v.35, p.898;
\par
M. Born, Einstein's Theory of Relativity, Dover, N.Y., 1962;
\par
C.M. Will, Intern. Journal of Modern Phys., 1992, v.1, p.13.
\end{document}